\documentclass[aps,prl,twocolumn,superscriptaddress]{revtex4-1}
\usepackage{graphics}
\usepackage[pdftex]{graphicx}
\usepackage{epsfig}
\usepackage{epsf}
\usepackage[leftFloats]{fltpage}
\usepackage{natbib}

\begin{document}
\bibliographystyle{plainnat}

\title{Network Physiology reveals relations between\\ network topology and physiological function}

\author{Amir Bashan}
\affiliation{Department of Physics, Bar-Ilan University, Ramat Gan, 52900 Israel.}
\author{Ronny P. Bartsch}
\affiliation{Harvard Medical School and Division of Sleep Medicine, Brigham and Women’s Hospital, Boston, MA 02115, USA.}
\author{Jan W. Kantelhardt}
\affiliation{Institute of Physics, Martin-Luther-Universit\"at Halle-Wittenberg, D-06099 Halle (Saale), Germany.}
\author{Shlomo Havlin}
\affiliation{Department of Physics, Bar-Ilan University, Ramat Gan, 52900 Israel.}
\author{Plamen Ch. Ivanov}
\affiliation{Harvard Medical School and Division of Sleep Medicine, Brigham and Women’s Hospital, Boston, MA 02115, USA.}
\affiliation{Department of Physics and Center for Polymer Studies, Boston University, Boston, MA}
\affiliation{Institute of Solid State Physics, Bulgarian Academy of Sciences, Sofia 1784, Bulgaria}


\begin{abstract}
The human organism is an integrated network where complex physiologic systems, 
each with its own regulatory mechanisms, continuously interact, and where failure 
of one system can trigger a breakdown of the entire network. Identifying and 
quantifying dynamical networks of diverse systems with different types of 
interactions is a challenge. Here, we develop a framework to probe interactions 
among diverse systems, and we identify a physiologic network. We find that each 
physiologic state is characterized by a specific network structure, demonstrating 
a robust interplay between network topology and function. Across 
physiologic states the network undergoes topological transitions associated 
with fast reorganization of physiologic interactions on time scales of a few minutes, 
indicating high network flexibility in response to perturbations. The proposed system-wide 
integrative approach may facilitate the development of a new field, Network Physiology.
\end{abstract}

\maketitle



Physiologic systems under neural regulation exhibit
high degree of complexity with nonstationary, intermittent, scale-
invariant and nonlinear behaviors \citep{Bassingthwaighte-1994,West-Science-1997}. Moreover, physiologic
dynamics transiently change in time under different physiologic
states and pathologic conditions \citep{Bunde-PRL-2000,%
Dvir2002,Ivanov-Nature-1996}, in response to
changes in the underlying control mechanisms. This complexity is further compounded by various
coupling \citep{Schaefer-Nature-1998,Bartsch-PRL-2007} and feedback 
interactions \citep{Guyton1955,Ivanov-EPL-1998,Hegger-PRL-1998} among different systems, the
nature of which is not well-understood. Quantifying these physiologic 
interactions is a challenge as one system may exhibit multiple
simultaneous interactions with other systems where the strength
of the couplings may vary in time. 
To identify the network of interactions between integrated physiologic systems, and to study 
the dynamical evolution of this network in relation to different physiologic states, it is necessary
to develop methods that quantify interactions between diverse systems.

Recent studies have identified networks with complex topologies \citep{Albert-Review-2002,Song-Nature-2005,Dorogovtsev-Review-2002}, 
have focused on emergence of self-organization and complex network behavior
out of simple interactions \citep{Williams-2000,Zhou-PRL-2006,Hasty-Nature-2002,Bhalla-1999},
on network robustness \citep{Newman2003,Pastor-Satorras2001,Banavar1999}, and more recently on critical
transitions due to failure in the coupling of interdependent networks \citep{Buldyrev-Nature-2010}. 
Growth dynamics of structural networks have been investigated in network models \citep{Albert-Review-2002,Dorogovtsev-Review-2002}, 
and in physical systems \citep{Dorogovtsev-Review-2002,Boccaletti-Review-2006}, and various structural and functional 
brain networks have been explored \citep{Boccaletti-Review-2006,Bullmore-Review-2009}. 
However, understanding the relation between topology and dynamics of 
complex networks remains a challenge, especially when networks are comprised of diverse systems with 
different types of interaction, each network node represents a multicomponent complex system
with its own regulatory mechanism, the output of which can vary in time,
and when transient output dynamics of individual nodes affect
the entire network by reinforcing (or weakening) the links and
changing network topology. A prime example of a combination of 
all these network characteristics is the human organism, where 
integrated physiologic systems form a network of interactions 
that affects physiologic function, and where breakdown in 
physiologic interactions may lead to a cascade of
system failures \citep{Buchman-2006}.\\\\
\begin{figure*}
\centering
\includegraphics[scale=0.42]{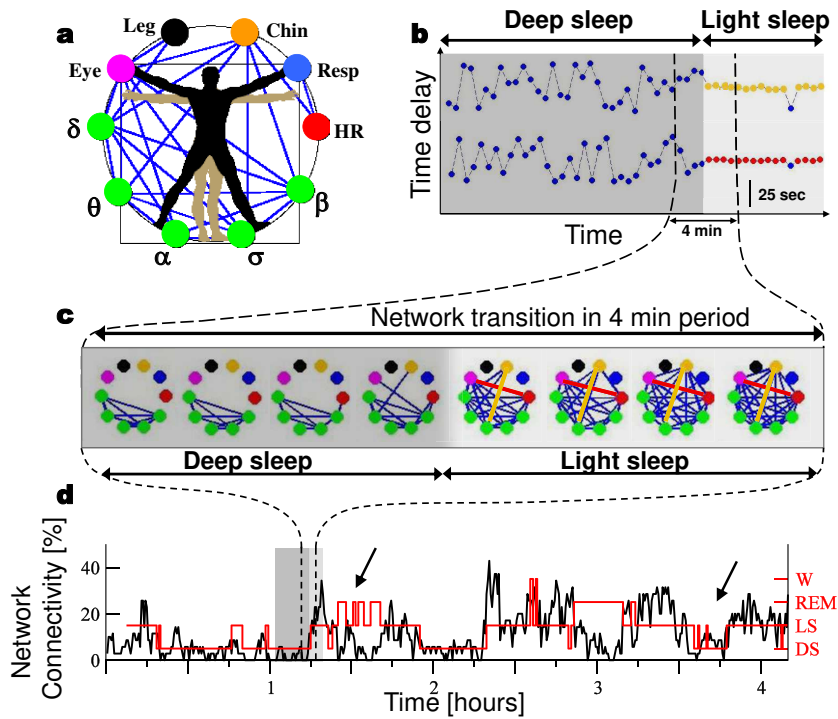}
\caption{{\bf Transitions in the network of physiologic interactions.} 
{\bf (a)} Dynamical network of interactions between physiological systems where ten 
network nodes represent six physiologic systems -- brain activity 
(EEG waves: $\delta$, $\theta$, $\alpha$, $\sigma$, $\beta$), cardiac 
(HR), respiratory (Resp), chin muscle tone, leg and eye movements. 
{\bf (b)} Transition in the interactions between physiologic systems across sleep stages.
The time delay between two pairs of signals, (top) $\alpha$-brain 
waves and chin muscle tone, and (bottom) HR and eye movement, quantifies their
physiologic interaction: highly irregular behavior (blue dots) during deep
sleep is followed by a period of time delay stability 
during light sleep indicating a stable physiologic interaction 
(red dots for the HR-eye and orange dots for the $\alpha$-chin interaction).
{\bf (c)} Transitions between physiologic states are associated with changes in network topology: 
snapshots over 30-sec windows during a transition from deep sleep (dark
gray) to light sleep (light gray). During deep sleep the network consists mainly of 
brain-brain links. With transition to light sleep links between other physiologic systems (network nodes) emerge
and the network becomes highly connected. The stable $\alpha$-chin and HR-eye interactions during light sleep 
in {\bf (b)} are shown by an orange and a red network link respectively.
{\bf (d)} Physiologic network connectivity for one subject during night sleep calculated in 30-sec windows
as the fraction (\%) of present links out of all possible links.
(brain-brain links not included, see Fig. \ref{Histograms}e). Red line marks sleep stages as scored in a sleep lab.
Low connectivity is consistently observed during deep sleep (0:30--1:15h and 1:50--2:20h) 
and REM sleep (1:30--1:45h and 2:50--3:10h), while transitions to light sleep and wake 
are associated with a significant increase in connectivity.}
\label{DynamicalNetwork}
\end{figure*}
We investigate the network of interactions between physiologic systems, and we focus on
the topology and dynamics of this network
and their relevance to physiologic function. We hypothesize
that during a given physiologic state the physiologic network
may be characterized by a specific topology and coupling strength between systems.
Further, we hypothesize that
coupling strength and network topology may abruptly change in response to transition from
one physiologic state to another. 
Such transitions may also be associated with changes in the connectivity of specific network nodes,
i.e., the number of systems to which a given physiologic system is
connected can change, forming sub-networks of physiologic interactions.
Probing physiologic network connectivity and the stability 
of physiologic coupling across physiologic states may thus provide new insights on 
integrated physiologic function. Such a systems-wide perspective on physiologic interactions,
tracking multiple components simultaneously, is necessary to understand the relation between
network topology and function.

\begin{figure*}
\centering
\includegraphics[scale=0.57]{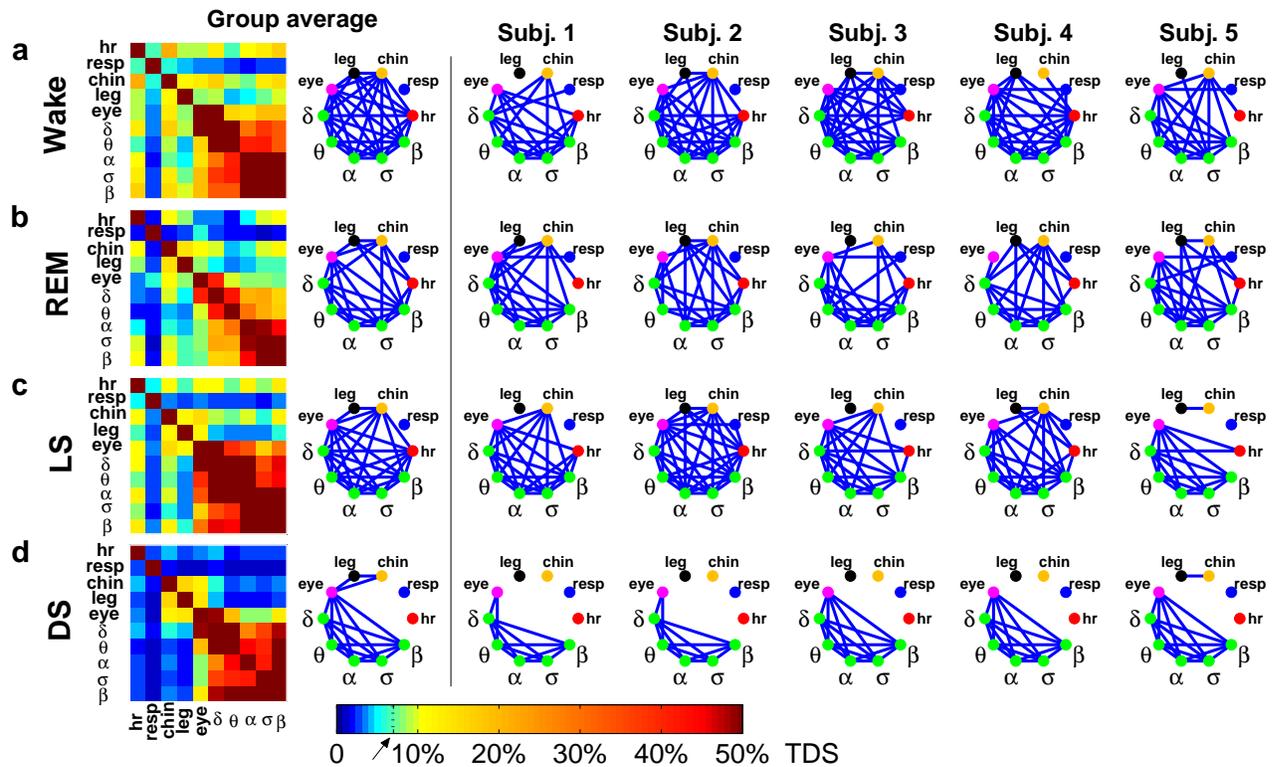}
\caption{{\bf Network connectivity across sleep stages.}
Group-averaged time delay stability (TDS) matrices and related networks of physiological 
interactions during different sleep stages: {\bf (a)} wake; {\bf (b)} REM sleep; {\bf (c)} light sleep (LS); {\bf (d)} deep sleep (DS). 
Matrix elements are obtained by quantifying the TDS for each pair of physiologic systems after obtaining the weighted average
of all subjects in the group: \% TDS = ($\sum_i s_i$/$\sum_i L_i$)$\times 100$ where $L_i$ indicates the total duration of a given sleep stage
for subject $i$, and $s_i$ is the total duration of time delay stability within $L_i$ for the considered pair of physiologic signals.
Color code represents the average strength of interaction between systems quantified as the 
fraction of time (out of the total duration of a given sleep-stage throughout 
the night) when TDS is observed. A network link between two systems
is defined when their interaction is characterized by a TDS of $\ge$ 7\% (arrow), a threshold determined by
surrogate analysis (see Methods). The physiologic network exhibits transitions across sleep stages
--- lowest number of links during deep sleep {\bf (d)}, higher during REM {\bf (b)} and highest during light sleep {\bf (c)}
and quiet wake {\bf (a)} --- a behavior observed in the group-averaged network as well as 
for each subject. Network topology also changes with sleep-stage 
transitions: from predominantly brain-brain links during deep sleep to a high number of brain-periphery and
periphery-periphery links during light sleep and wake.}
\label{TDS-Matrices}
\end{figure*}

\section*{Results}

\subsection*{Time delay stability and network of physiologic interactions}

\begin{figure}
\centering
\includegraphics[scale=0.5]{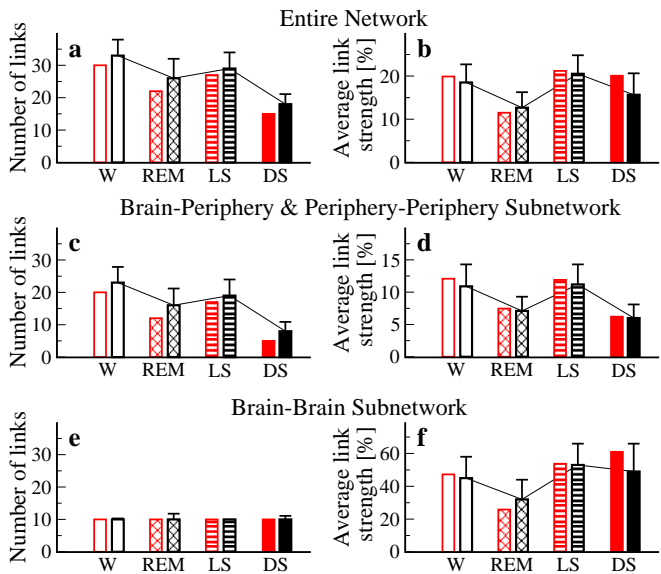}
\caption{{\bf Sleep-stage stratification pattern in network connectivity and network link strength.}
Group-averaged number of links {\bf (a)} and averaged link strength {\bf (b)} are
significantly higher during wake and light sleep compared to REM and deep sleep (student
t-test $p<10^{-3}$ for both quantities when comparing REM and deep sleep with wake and light sleep). There is
no significant difference between wake and light sleep ($p>5\times10^{-2}$). This pattern is even more 
pronounced for the subnetwork formed by the brain-periphery and periphery-periphery links
shown in {\bf (c)} and {\bf (d)} ($p<10^{-6}$ for both quantities when comparing REM and deep sleep with 
wake and light sleep). In contrast, the number of brain-brain links remains practically unchanged
with sleep-stage transitions {\bf (e)}, and the average brain-brain link is $\approx$ 5 times 
stronger in all sleep stages compared to the other network links {\bf (f)}. 
The group-averaged patterns in the number of network links and in the average link strength 
across sleep stages (black bars) are consistent with the behavior observed for individual
subjects (red bars in all panels represent the same subject).
The group-averaged number of links for each sleep stage is obtained from the corresponding
group-averaged network in Fig. \ref{TDS-Matrices}.
The average link strength is measured in \% TDS and is obtained by taking the mean of
all elements in the TDS matrix for each sleep stage (Fig. \ref{TDS-Matrices}); it represents the average
strength of all links in a network obtained from a given subject during a specific sleep stage 
which then is averaged over all subjects. Error bars indicate standard deviation obtained from 
a group of 36 subjects (Methods).}
\label{Histograms}
\end{figure}

\begin{figure*}
\centering
{\includegraphics[scale=0.7]{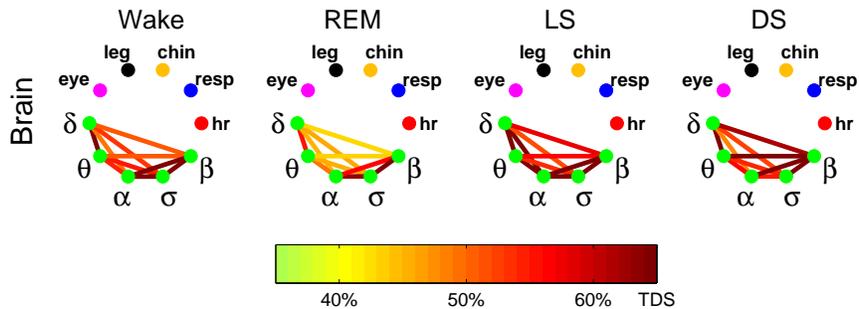}}
\caption{{\bf Network connectivity and link strength of the brain-brain subnetwork
for different sleep stages.} While the topology of the brain subnetwork does not change,
the strength of network links significantly changes with strongest links during light sleep and deep sleep (brown and dark red
color), intermediate during wake (red and orange color) and weakest links during REM sleep (yellow color).}
\label{Brain}
\end{figure*}

\begin{figure*}
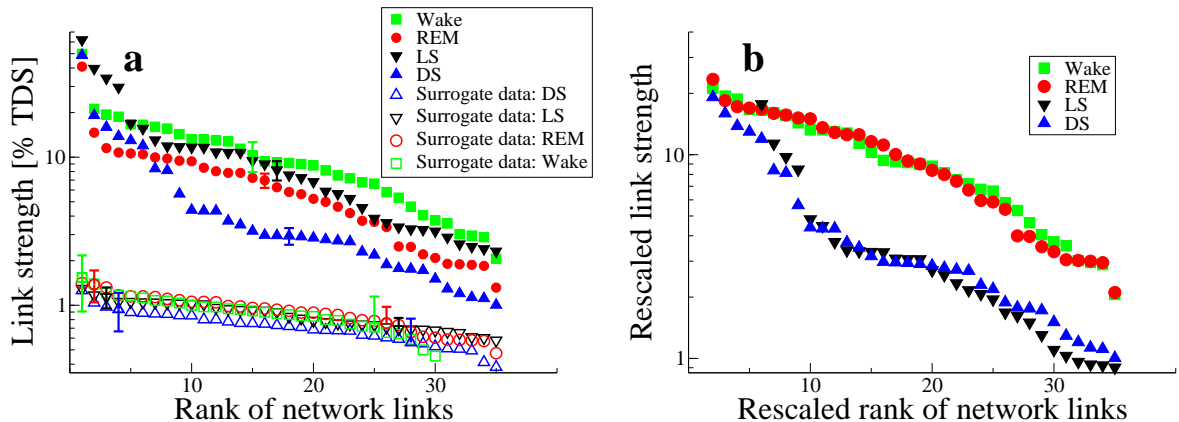

\centering
\hspace{-1cm}
{\includegraphics[scale=0.3]{Rank-TDS-loglin.eps} \hspace{0.25cm}
\includegraphics[scale=0.3]{Rank-TDS-aligned-loglin.eps}}
\caption{{\bf Rank distributions of the strength of network links.}
Group-averaged strength of individual physiologic network links for different sleep
stages. Rank 1 corresponds to the strongest link in the network, i.e., highest degree of 
time delay stability (TDS) (shown are all periphery-periphery and brain-periphery links). 
{\bf (a)} The rank distributions for different sleep stages are characterized by different strength of the network links measured in \% TDS ---
consistently lower values for most links during deep sleep, higher values during REM and highest during light sleep and wake,
indicating that the stratification pattern in Fig. \ref{Histograms}d is present not only for the average link 
strength (when averaging over different types of links in the network) but also for 
the strength of individual links. Indeed, links from all ranks are consistently
stronger in light sleep compared to deep sleep and REM: such rank-by-rank comparison of links across sleep stages is possible 
because the rank order of the links does not change significantly from one sleep stage to another 
(Wilcoxon signed-rank test for all pairs of rank distributions yields $0.77\le p\le 0.93$). 
A surrogate test based on TDS analysis of signals paired from different subjects, which eliminates
endogenous physiologic coupling, leads to significantly reduced
link strength ($p<10^{-3}$) and close to uniform rank distributions with no difference between sleep stages (open symbols), 
indicating that the TDS method uncovers physiologically-relevant information. Error bars for the original and surrogate 
data indicate the standard error for a specific link when averaged over all 36 subjects or over 36 surrogate pairs respectively.
{\bf (b)} Rescaling the plots reveals two distinct forms of rank distributions: a slow decaying distribution
for wake and REM, and a fast decaying distribution for light sleep and deep sleep with a pronounced plateau in the middle rank range corresponding to a
cluster of links with similar strength, most of which related to the cardiac system.}
\label{RankPlots}
\end{figure*}

\begin{figure*}
\centering
{\includegraphics[scale=0.65]{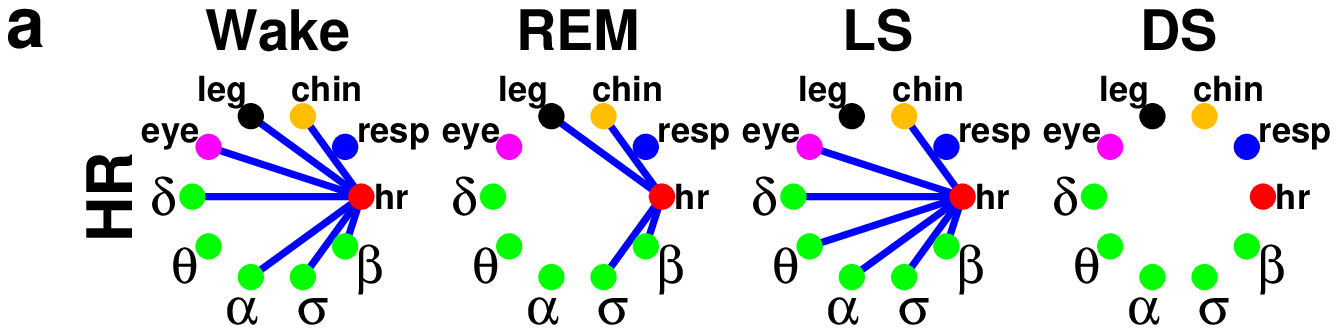}\includegraphics[scale=0.55]{figure5_HR_AvgLinksStrength-resc.eps}}\\
{\includegraphics[scale=0.65]{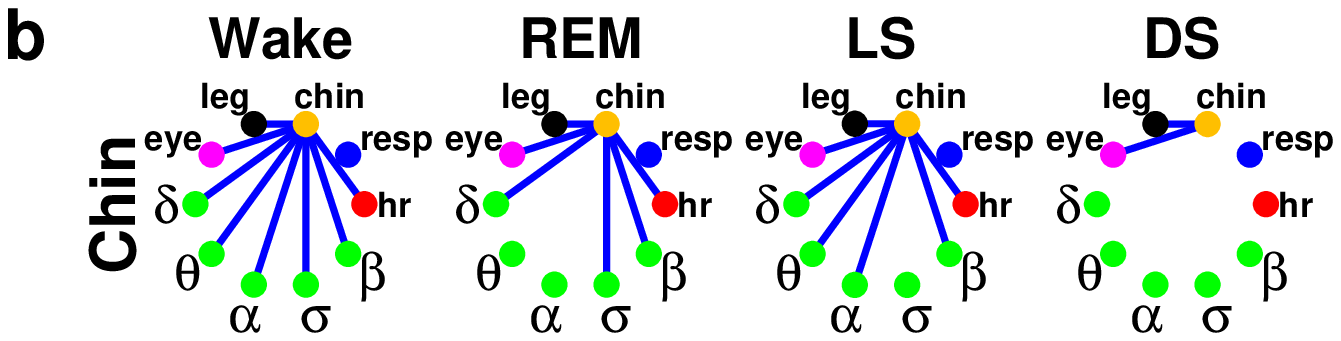}\includegraphics[scale=0.55]{figure5_Chin_AvgLinksStrength-resc.eps}}\\
\caption{{\bf Transitions in connectivity and link strength of 
individual network nodes across sleep stages.} 
The number of links to specific network nodes significantly changes, 
with practically no links during deep sleep, a few links during REM and much higher connectivity during
light sleep and wake. Notably, the average strength of the links connecting a given network node is also lowest during deep sleep and highest during 
light sleep and wake. Shown are connectivity and average link strength for two network nodes: {\bf (a)} heart and {\bf (b)} chin. This sleep-stage
stratification pattern in individual node connectivity and in the average strength of the links connecting a specific
network node is consistent with the transitions of the entire network across sleep stages shown in Fig. \ref{Histograms} c and d.
Networks for {\bf (a)} heart and {\bf (b)} chin are obtained by averaging the corresponding networks for all subjects. During deep sleep no links to the heart 
are shown since the strength of each link averaged over all subjects is below the significance threshold (Fig. \ref{TDS-Matrices}
and Fig. \ref{threshold-def}, Methods). Right bars in the panels represent for different sleep stages the group mean of the average strength 
of network links connecting heart and chin respectively, and error bars show the standard deviation. 
Left bars represent an individual subject. Note that the absence of a link between heart rate and respiration in the physiologic network does not 
indicate absence of cardio-respiratory coupling but rather that this coupling as represented by time delay stability (TDS) is rarely
stable for periods longer than 2--4 min (where 2 min is the minimum window over which TDS is determined; Method section),
and that cardio-respiratory TDS episodes form less than 7\% of the recordings, which is the significance threshold level (Method section).
Such ``on'' and ``off'' intermittent interaction between these two systems is observed also in other independent measures 
of cardio-respiratory coupling --- respiratory sinus arrhythmia (RSA) \citep{Angelone-RSA-1964,Song-RSA-2003} and the degree 
of phase synchronization \citep{Schaefer-Nature-1998} --- where relatively short ``on'' episodes are separated by periods 
of no interrelation as quantified by these measures.}
\label{Hubs}
\end{figure*}

The framework we propose is based on a complex networks approach to quantify physiologic
interactions between diverse physiologic systems, where network nodes represent different physiologic
systems and network links indicate the dynamical interaction (coupling) between systems. This framework
allows to quantify the topology and the associated dynamics in the links strength of physiologic 
networks during a given physiologic state, taking into account the signal output
of individual physiologic systems as well as the interactions among them, and to track
the evolution of multiple interconnected systems undergoing transitions from one
physiologic state to another (Fig. \ref{DynamicalNetwork}). We introduce the concept of time delay stability (TDS) to identify and 
quantify dynamic links among physiologic systems.
We study the network of interactions for an ensemble of key integrated physiologic systems 
(cerebral, cardiac, respiratory, ocular and muscle activity). We consider different sleep stages (deep, 
light, rapid eye movement (REM) sleep and quiet wake) as examples of physiologic states.
While earlier studies have identified how sleep regulation influences aspects of the specific control mechanism 
of individual physiologic systems (e.g., cardiac or respiratory \citep{Bunde-PRL-2000,Dvir2002,Otzenberger1998,%
Schumann-Sleep-2010}) or have focused on the organization of functional 
connectivity of EEG networks during sleep \citep{Ferri2008} and under neurological disorders such 
as epilepsy \citep{Schindler2008}, the dynamics and topology of a physiologic network comprised of diverse systems 
have not been studied so far. Further, the relation between network topology and function, and how it changes with transitions 
across distinct physiologic states is not known.
We demonstrate that sleep stages are associated with markedly
different networks of physiologic interactions (Fig. \ref{TDS-Matrices}) characterized by different
number and strength of links (Figs. \ref{Histograms} and \ref{Brain}), by different 
rank distributions (Fig. \ref{RankPlots}), and by specific node connectivity (Fig. \ref{Hubs}). 
Traditionally, differences between sleep stages are attributed to modulation in the sympatho-vagal balance with
dominant sympathetic tone during wake and REM \citep{Otzenberger1998}:
spectral, scale-invariant and nonlinear characteristics of 
the dynamics of individual physiologic systems indicate higher 
degree of temporal correlations and nonlinearity during wake and 
REM compared to non-REM (light and deep sleep) where physiologic dynamics exhibit weaker correlations and loss of
nonlinearity \citep{Bunde-PRL-2000, Schumann-Sleep-2010}. In contrast, the network of 
physiologic interactions shows a completely different picture: the network characteristics during 
light sleep are much closer to those during wake and very different from deep sleep (Figs. \ref{TDS-Matrices}
and \ref{Histograms}). Specifically, we find that network connectivity and
overall strength of physiologic interactions are significantly higher during
wake and light sleep, intermediate during REM and much 
lower during deep sleep. Thus, our empirical observations
indicate that while sleep-stage related modulation in sympatho-vagal 
balance plays a key role in regulating individual physiologic systems,
it does not account for the physiologic network topology and dynamics across 
sleep stages, showing that the proposed framework captures principally new information.

\newpage
To quantify the interaction between physiologic systems and to probe 
how this interaction changes in time under different physiologic conditions 
we study the time delay with which modulations in
the output dynamics of a given physiologic system are consistently followed by 
corresponding modulations in the signal output of another system. 
Periods of time with approximately constant time delay indicate a stable physiologic
interaction, and stronger coupling between physiologic systems results
in longer periods of time delay stability (TDS). 
Utilizing the TDS method we build a dynamical network 
of physiologic interactions, where network links between physiological systems
(considered as network nodes) are established when the time delay stability 
representing the coupling of these systems exceeds a significance threshold level,
and where the strength of the links is proportional to the percentage of time for which
time delay stability is observed (Methods). 

\subsection*{Transitions in network topology with physiologic function}
We apply this new approach to a group of healthy young subjects (Methods). 
We find that the network of interactions between physiologic systems is very sensitive to 
sleep-stage transitions. In a short time window of just a few minutes the network
topology can dramatically change --- from only
a few links to a multitude of links (Fig. \ref{DynamicalNetwork}) --- indicating transitions in the
global interconnectivity between physiological systems.
These network transitions are not associated with random occurrence or loss of 
links but are characterized by certain organization in network topology
where given links between physiological systems remain stable during the transition
while others do not --- e.g., brain-brain links persist during the transition from deep sleep to 
light sleep while brain-periphery links significantly change (Fig. \ref{DynamicalNetwork}c). 
Further, we find that sleep-stage transitions are paralleled by abrupt
jumps in the total number of links leading to higher or lower network connectivity (Fig. \ref{DynamicalNetwork}c, d).      
These network dynamics are observed for each subject in the database, where
consecutive episodes of sleep stages are paralleled by a level of connectivity 
specific for each sleep stage, and where sleep-stage transitions
are consistently followed by transitions in network connectivity throughout the
course of the night (Fig. \ref{DynamicalNetwork}d). Indeed, the network of physiologic
interactions exhibits a remarkable responsiveness as network connectivity changes even for
short sleep-stage episodes (arrows in Fig. \ref{DynamicalNetwork}d), demonstrating
a robust relation between network topology and function. This is the first 
observation of a real network evolving in time and undergoing topological
transitions from one state to another. 

To identify the characteristic network topology for each sleep stage we obtain 
group-averaged time delay stability matrices, where each matrix element represents
the percentage of time with stable time delay between two physiological systems, 
estimated over all episodes of a given sleep stage throughout the
night. Matrix elements above a threshold of statistical significance
(Fig. \ref{threshold-def}, Methods) indicate stable interactions
between physiologic systems represented by network links (Fig. \ref{TDS-Matrices}). We find that
matrix elements greatly vary for different sleep stages with much higher values
for wake and light sleep, lower values for REM and lowest for deep sleep. This
is reflected in higher network connectivity for wake and light sleep, lower for REM and significantly reduced number of links 
during deep sleep (Fig. \ref{Histograms}a). Further, the TDS matrices indicate separate
subgroups of interactions between physiologic systems --- brain-periphery, periphery-periphery
and brain-brain interactions --- that are affected differently during
sleep stages and form different sub-networks. Specifically, matrix elements
representing interactions between peripheral systems (cardiac, respiratory,
chin, eye, leg) and the brain as well as interactions among the peripheral
systems are very sensitive to sleep-stage transitions, leading to different network topology 
for different sleep stages (Fig. \ref{TDS-Matrices}). We find sub-networks with high number 
of brain-periphery and periphery-periphery links during wake and light sleep, lower number of links during REM 
and a significant reduction of links at deep sleep (Fig. \ref{Histograms}c). In contrast, matrix
elements representing brain-brain interactions form a subnetwork 
with the same number of brain-brain links (Fig. \ref{Histograms}e), and stable topology
consistently present in the physiologic network during all sleep stages (Fig. \ref{TDS-Matrices}).
Sleep-stage related transitions in network connectivity and topology
are not only present in the group-averaged data but also in the physiologic 
networks of individual subjects, suggesting universal behavior (Fig. \ref{TDS-Matrices}). 
Notably, we find a higher number of brain-periphery links during REM compared to
deep sleep despite inhibition of motoneurons in the brain leading to muscle 
atonia during REM \citep{Dement-Book-1994}. The empirical observations of 
significant difference in network connectivity and topology during light sleep compared to 
deep sleep are surprising, given the similarity in spectral, scale-invariant and nonlinear
properties of physiologic dynamics during light sleep and deep sleep \citep{Bunde-PRL-2000,Dvir2002,Otzenberger1998,%
Schumann-Sleep-2010} (both stages traditionally classified
as non-rapid eye movement sleep (NREM)), and indicate that previously unrecognized aspects 
of sleep regulation may be involved in the control of physiologic network interactions. 

\subsection*{Physiologic states and network link strength stratification}
Networks with identical connectivity and topology can exhibit different strength
of their links. Network link strength is determined as the fraction of time when
TDS is observed (Methods). We find that the
average strength of network links changes with sleep-stage transitions:
network links are significantly stronger during wake and light sleep compared to
REM and deep sleep --- a pattern similar to the behavior of the network connectivity
across sleep stages (Fig. \ref{Histograms}a, b). Further, subnetworks of physiologic interactions
exhibit different relationship between connectivity and average link strength. 
Specifically, the subnetwork of brain-periphery and
periphery-periphery interactions is characterized by significantly stronger
links (and also higher connectivity) during wake and light sleep, and much
weaker links (with lower network connectivity) during deep sleep and REM (Fig. \ref{Histograms}c, d).
In contrast, the subnetwork of brain-brain interactions exhibits very
different patterns for the connectivity and the average link strength --- while the
group average subnetwork connectivity remains constant across sleep stages,
the average link strength varies with highest values during light sleep and deep sleep,
and a dramatic $\approx 40\%$ decline during REM. The observation of
significantly stronger links in the brain-brain subnetwork during NREM
compared to REM sleep is consistent with the characteristic of NREM as 
EEG-synchronized sleep and REM as EEG-desynchronized sleep \citep{Dement-Book-1994}.
During NREM sleep adjacent cortical neurons fire synchronously with a relatively
low frequency rhythm \citep{Siegel-Nature-2005} leading to coherence between
frequency bands in the EEG signal, and thus to stable time delays and strong
network links (Fig. \ref{Histograms}f). In contrast, during REM sleep cortical neurons are highly active but 
fire asynchronously \citep{Siegel-Nature-2005}, resulting in weaker links (Fig. \ref{Histograms}f).
Our findings of stronger links in the brain-brain subnetwork during non-REM sleep (Fig. \ref{Histograms}f and Fig. \ref{Brain}) indicate that
bursts (periods of sudden temporal increase) in the spectral power of one EEG-frequency band are consistently synchronized in
time with bursts in a different EEG-frequency band, thus leading to longer periods of time delay stability and correspondingly
stronger network links.
This can explain some seemingly surprising
network links --- for example, we find a strong link between $\alpha$ and 
$\delta$ brain activity during non-REM sleep (Fig. \ref{TDS-Matrices})
although $\alpha$ waves are greatly diminished and $\delta$ waves are dominant \citep{Dement-Book-1994}. Since the spectral
densities of both waves are normalized before the TDS analysis (Methods),
the presence of a stable $\alpha$ - $\delta$ link indicates that a relative
increase in the spectral density in one wave is followed, with a stable
time delay, by a corresponding increase in the density of the other wave ---
an intriguing physiologic interaction which persists not only during deep sleep
but is also present in light sleep, REM and quiet wake (Fig. \ref{TDS-Matrices}).
Notably, the average link strength of the brain-brain subnetwork is
by a factor of $\approx 5$ higher compared to all other links in the physiologic 
network (Fig. \ref{Histograms}d, f). 

The finding of completely different sleep-stage stratification patterns in key network properties of the brain-brain 
subnetwork compared to the periphery-periphery/brain-periphery subnetworks suggests a very 
different role these sub-networks play in coordinating physiologic interactions during sleep.
The similarity in the brain-brain subnetwork during deep sleep and light sleep indicates that the proposed 
TDS approach is sensitive to quantify synchronous slow-wave brain activity during NREM sleep
that leads to stronger brain-brain links during light sleep and deep sleep ($\approx$ 50-60\% TDS) compared 
to REM ($\approx$ 35\% TDS), as shown in (Fig. \ref{Histograms}f and Fig. \ref{Brain}). 
The significant difference between light sleep and deep sleep observed for the periphery-periphery/brain-periphery 
subnetwork in the number of links (t-test: $p<10^{-12}$) as well as in the average link strength 
(t-test: $p<10^{-11}$), indicates that the interactions between physiologic dynamics outside the 
brain are very different during these sleep stages.

Our finding that the average link strength 
exhibits a specific stratification pattern across sleep stages (Fig. \ref{Histograms}) 
raises the question whether the underlying distribution of the network links
strength is also sleep-stage dependent. To this end we probe the relative strength
of individual links, and we obtain the rank distribution of the strength of 
network links for each sleep stage averaged over 
all subjects in the group (Fig. \ref{RankPlots}a). We find that the rank distribution corresponding 
to deep sleep is vertically shifted to much lower values for the strength of
the network links, while the rank distribution for light sleep and wake is for all links
consistently higher than the distribution for REM. Thus, the sleep-stage stratification 
pattern we find for the average strength of the network links (Fig. \ref{Histograms}d)
originates from the systematic change in the strength of individual network links with
sleep-stage transitions. Notably, while 
the strength of individual network links changes significantly with 
sleep stages, the rank order of the links does not significantly change.    
After rescaling the rank distributions for light sleep and REM (by
horizontal and vertical shifts), we find that they collapse onto the rank plots
of deep sleep and wake respectively, following two distinct functional forms:
a slow and smoothly decaying rank distribution for REM and wake, and 
a much faster decaying rank distribution for deep sleep and light sleep
with a characteristic plateau in the mid rank range indicating a cluster 
of links with similar strength (Fig. \ref{RankPlots}b). We note that, although
the form of the rank distributions for deep sleep and light sleep as well as for wake and REM are respectively
very similar, the average strength of the links is significantly different 
between deep sleep and light sleep and between wake and REM (Fig. \ref{Histograms}d).

\subsection*{Local topology and connectivity of the physiologic network}
Our observations that physiologic networks undergo dynamic transitions where key global
properties significantly change with sleep-stage transitions, raise the 
question whether local topology and connectivity of individual network nodes also
change during these transitions. Considering each physiologic system (network
node) separately, we examine the number and strength of all links connecting
the system with the rest of the network. Specifically, we find that the cardiac
system is highly connected to other physiologic systems in the network during 
wake and light sleep (Fig. \ref{Hubs}). In contrast, during deep sleep we do not find 
statistically significant time delay stability in the interactions of the cardiac 
system, which is reflected by absence of cardiac links (Fig. \ref{Hubs}). Further,
we find that the average strength of the links connected to the cardiac system
also changes with sleep stages: stronger interactions (high \% TDS) during wake and
light sleep, and significantly weaker interactions below the significance threshold during 
deep sleep (Fig. \ref{Hubs}). Such `isolation' of the cardiac node from the rest
of the network indicates a more autonomous cardiac function during deep sleep ---
also supported by earlier observations of breakdown of long-range correlations and 
close to random behavior in heartbeat intervals in this sleep stage \citep{Bunde-PRL-2000}. 
Transition to light sleep, REM and wake, where the average link strength
and connectivity of the cardiac system is significantly higher indicating increased 
interactions with the rest of the network, leads to correspondingly higher degree  
of correlations in cardiac dynamics \citep{Bunde-PRL-2000}. 
Similarly, respiratory dynamics also exhibit high degree of correlations during
REM and wake, lower during light sleep and close to random behavior during
deep sleep \citep{Schumann-Sleep-2010}. 
Such transitions in the number and strength of links across sleep
stages we also find for other network nodes (Fig. \ref{Hubs}).
Moreover, the sleep-stage stratification pattern in connectivity and
average link strength for individual network nodes (Fig. \ref{Hubs}) is consistent
with the pattern we observe for the entire network (Fig. \ref{Histograms}).
Our findings of significant reduction in the number and strength of brain-periphery and 
periphery-periphery links in the corresponding sub-networks during deep sleep 
indicate that breakdown of cortical interactions, previously reported during deep sleep \citep{Massimini-Science-2005}, may
also extend to other physiologic systems under neural regulation. Indeed, the low connectivity in the 
physiologic network we find in deep sleep may explain why people awakened during deep sleep do not adjust immediately 
and often feel groggy and disoriented for a few minutes. This effect is not observed if subjects are awakened from 
light sleep \citep{Dement-Book-1994} when we find the physiologic network to be highly connected (Fig. \ref{TDS-Matrices}). 
Further, since risk of predation modifies sleep architecture \citep{Lima2005,Lesku2006,Capellini2008} and since abrupt awakening
from deep sleep is associated with increased sleep inertia, higher sensory threshold, 
and impaired sensory reaction and performance \citep{Downey1987,Tassi2000}
that may lead to increased vulnerability, the fact that deep sleep (lowest physiologic network connectivity) dominates at the 
beginning of the night and not close to dawn, when many large predators preferably hunt, may have been evolutionarily advantageous.

\section*{Discussion}
Introducing a framework based on the concept of TDS we identify
a robust network of interactions between physiologic systems, 
which remains stable across subjects during a given physiologic state.
Further, changes in the physiologic state lead to complex network transitions 
associated with a remarkably structured reorganization of network connectivity 
and topology that simultaneously occurs in the entire
network as well as at the level of individual network nodes, while
preserving the hierarchical order in the strength of individual network
links. Such network transitions lead to the formation of sub-networks of physiologic
interactions with different topology and dynamical characteristics. 
In the context of sleep stages, network transitions are characterized by
a specific stratification pattern where network connectivity and link strength are
significantly higher during light sleep compared to deep sleep and during
wake compared to REM. This can not be explained by the dynamical characteristics of the output 
signals from individual physiologic systems which are similar during light sleep and deep sleep 
as well as during wake and REM. The dramatic change in network structure with transition
from one physiologic state to another within a short time window indicates a high 
flexibility in the interaction between physiologic systems in response to change in physiologic
regulation. Such change in network structure in response to change in the mechanisms of control 
during different physiologic states suggests that our findings reflect intrinsic features of 
physiologic interaction. The observed stability in network topology and
rank order of links strength during sleep stages, and
the transitions in network organization across sleep stages provide new insight 
into the role which individual physiologic systems as well as their interactions 
play during specific physiologic states. 
While our study is limited to a data-driven approach these empirical findings may facilitate future efforts on developing 
and testing network models of physiologic interaction.
This system-wide integrative approach to individual systems and the network of their interactions may facilitate the emergence of
a new dimension to the field of systems physiology \citep{Guyton1955} that will include not only interactions within but also across physiologic systems.
In relation to critical clinical care, where multiple organ failure is often the reason for fatal outcome \citep{Buchman-2006,Deitch-1992}, 
our framework may have practical utility in assessing whether dynamical links between physiologic systems remain substantially
altered even when the function of specific systems is restored after treatment \citep{Lizana2000}.
While we demonstrate one specific application,
the framework we develop can be applied to a broad range of complex systems where the TDS method
can serve as a tool to characterize and understand the dynamics and function of real-world heterogeneous and
interdependent networks. The established relation between dynamical network topology and 
network function has not only significant medical and clinical implications, but is also of 
relevance for the general theory of complex networks. 

\section*{Methods}


\subsection*{Data}
We analyze continuously recorded multi-channel physiologic data obtained from 36 healthy 
young subjects (18 female, 18 male, with ages between 20-40, average 29 years) during 
night-time sleep \citep{Kloesch-IEEE-2001} (average record duration is 7.8 hours). This allows 
us to track the dynamics and evolution of the network of physiologic interactions during different sleep stages
and sleep-stage transitions (Fig. \ref{DynamicalNetwork}). We focus on physiologic dynamics during
sleep since sleep stages are well-defined physiological states, and external
influences due to physical activity or sensory inputs are reduced during sleep.
Sleep stages are scored in 30 sec epochs by sleep lab technicians based on standard
criteria. In particular, we focus on the electroencephalogram
(EEG), the electrocardiogram (ECG), respiration, the electrooculogram (EOG), and the electromyogram (EMG) of 
chin and leg. 
In order to compare these very different signals with each other and to study interrelations between them, we extract
the following time series from the raw signals: the spectral power of five frequency bands of the EEG in moving windows of 2 sec
with a 1 sec overlap: $\delta$ waves (0.5-3.5 Hz), $\theta$ waves (4-7.5 Hz), $\alpha$ waves (8-11.5 Hz), $\sigma$ waves (12-15.5 Hz),
$\beta$ waves (16-19.5 Hz); the variance of the EOG and EMG signals in moving windows of 2 sec with a 1 sec overlap; heartbeat RR-intervals
and interbreath intervals are both re-sampled to 1 Hz (1 sec bins) after which values are inverted to obtain heart rate and respiratory rate. 
Thus, all time series have the same time resolution of 1 sec before the TDS-analysis is applied.

\begin{figure}
\includegraphics[scale=0.3]{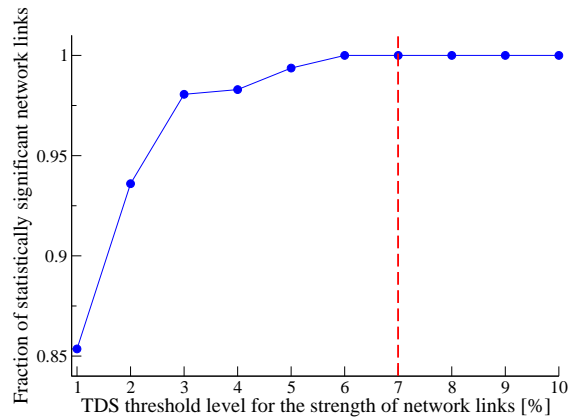}
\caption{{\bf Determining significance threshold for the strength of network links.}
With increasing the time delay stability (TDS) threshold level which allows only stronger links with higher TDS
values to be considered in the physiologic network, the fraction of statistically significant
network links that carry physiologically relevant information also increases, and at a 
significance threshold of $\approx$ 7\% TDS (marked by a vertical dashed line) all network
links (100\%) are statistically significant. Periphery-periphery and brain-periphery links
during all sleep stages are considered when determining this threshold. Statistical significance
of a specific physiologic link is estimated by comparing the strength distribution of this link across
all subjects in the group with a distribution of surrogate links representing ``interactions'' 
between the same systems paired from different subjects). Based on this surrogate test, a p-value 
$<10^{-3}$ obtained from the student t-test indicates statistically significant strength of a given link.}
\label{threshold-def}
\end{figure}

Utilizing sleep data as an example we demonstrate that a network approach 
to physiologic interactions is necessary to understand how modulations in the 
regulatory mechanism of individual systems translate into reorganization 
of physiologic interactions across the human organism.    

\begin{figure}
\centering
{\includegraphics[scale=0.3]{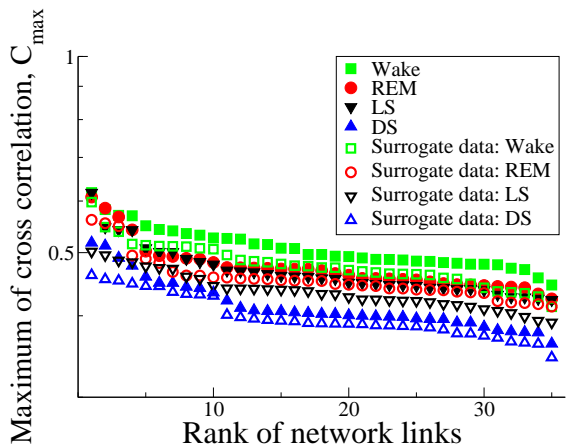}}
\caption{{\bf Cross-correlation and surrogate analysis.} Rank plots obtained from 
cross-correlation analysis show no statistically significant differences
between real and surrogate data, indicating that cross-correlation is not a reliable measure to identify physiologic 
interactions.}
\label{Rank-CCF}
\end{figure}

\subsection*{Time Delay Stability (TDS) Method}
Integrated physiologic systems are coupled by feedback and/or
feed forward loops with a broad range of time delays. 
To probe physiologic coupling we propose an approach based
on the concept of time delay stability: in the presence of stable/strong 
interactions between two systems, transient modulations in the 
output signal of one system lead to corresponding changes that
occur with a stable time lag in the output signal of another 
coupled system. Thus, long periods of constant time delay indicate
strong physiologic coupling.

The TDS method we developed for this study consists of the following steps:

(1.) To probe the interaction between two physiologic systems X and Y, we consider their output signals
$\{x\}$ and $\{y\}$ each of length $N$. We divide both signals $\{x\}$ and $\{y\}$ into $N_L$ overlapping segments $\nu$ 
of equal length $L=60$ sec. We choose an overlap of $L/2=30$ sec which corresponds
to the time resolution of the conventional sleep-stage scoring epochs, and thus $N_L=[2N/L]-1$.
Prior to the analysis, the signal in each segment $\nu$ is normalized separately to zero mean and unit standard deviation, in order 
to remove constant trends in the data and to obtain dimensionless signals. 
This normalization procedure assures that the estimated coupling between the signals $\{x\}$ and $\{y\}$ is not affected
by their relative amplitudes. 

(2.) Next, we calculate the cross-correlation function, 
$C^{\nu}_{xy}(\tau)=\frac{1}{L}\sum_{i=1}^{L} x^{\nu}_{i+(\nu -1){L\over 2}}~y^{\nu}_{i+(\nu -1){L\over 2}+\tau}$,
within each segment $\nu=1,\ldots,N_L$ by applying periodic boundary conditions. For each segment $\nu$
we define the time delay $\tau^{\nu}_0$ to correspond to the maximum in the absolute value 
of the cross-correlation function $C^{\nu}_{xy}(\tau)$ in this segment 
$\tau^{\nu}_0 = \tau \vert_{|C^\nu_{xy}(\tau)| \ge |C^\nu_{xy}(\tau')|~~\forall \tau'}$.
Time periods of stable interrelation between two signals are represented by segments of approximately
constant $\tau_0$ (light shade region in Fig. \ref{DynamicalNetwork}b) in the newly defined series of time delays, 
$\{\tau^{\nu}_0\}_{\nu=1,\ldots,N_L}$. In contrast, absence of stable coupling between the signals 
corresponds to large fluctuations in $\tau_0$ (dark shade region in Fig. \ref{DynamicalNetwork}b).

(3.) We identify two systems as linked if their corresponding signals exhibit a time delay that 
does not change by more than $\pm 1$ sec for several consecutive segments $\nu$. We track the values of 
$\tau_0$ along the series $\{\tau^{\nu}_0\}$: when for at least four out of five consecutive segments
$\nu$ (corresponding to a window of $5\times30$ sec) the time delay remains in the interval [$\tau_0-1$,$\tau_0+1$]
these segments are labeled as stable. This procedure is repeated for a sliding window with a step size one along the entire
series $\{\tau^{\nu}_0\}$. The \% TDS is finally calculated as the fraction of stable points 
in the time series $\{\tau^{\nu}_0\}$. 

Longer periods of TDS between the output signals of two systems reflect more stable
interaction/coupling between these systems. Thus, the strength of the links in the
physiologic network is determined by the percentage of time when TDS is observed:
higher percentage of TDS corresponds to stronger links. To identify physiologically
relevant interactions, represented as links in the physiologic network, we determine
a significance threshold level for the TDS based on comparison with surrogate data:
only interactions characterized by TDS values above the significance threshold are
considered. 

The TDS method is general, and can be applied to diverse systems. 
It is more reliable in identifying physiologic coupling compared
to traditional cross-correlation and cross-coherence analyses (Fig. \ref{Rank-CCF}) 
which are not suitable for heterogeneous and nonstationary signals, 
and are affected by the degree of auto-correlations in these 
signals~\citep{Podobnik-EPJB-2007}.

To compare interactions between physiologic systems which are very different
in strength and vary with change of physiologic state (e.g., transitions across sleep stages),
we define the significance threshold as the percent of TDS for which all links included 
in the physiologic network are statistically significant. To identify statistical significance
of a given link between two physiologic systems, we compare the distribution of TDS
values for this link obtained from all 36 subjects in our database with the distribution
of TDS values obtained for 100 surrogates of this link where the signal outputs from
the same two physiologic systems taken from different subjects are paired for the analysis
in order to eliminate the endogenous physiologic coupling. A student t-test was performed
to determine the statistical significance between the two distributions. This procedure
is repeated for all pairs of systems (links) in the network, and network links are 
identified as significant when the t-test p-value $<10^{-3}$. The significance threshold
level for TDS is then defined as the value above which all network links are statistically
significant, and thus represent endogenous interactions between physiologic systems.
We find that a threshold of approximately 7\% TDS is needed to identify networks of statistically 
significant links for all sleep stages (Fig. \ref{threshold-def}).

\subsection*{Surrogate tests}
To confirm that the TDS method captures physiologically relevant information about
the endogenous interactions between systems, we perform a surrogate test where
we pair physiologic signals from different subjects, thus eliminating physiologic coupling.
Applying the TDS method to these surrogate data, we obtain almost uniform rank distributions
with significantly decreased link strength (Fig. \ref{RankPlots}a) due to the absence of physiologic interactions.
Further, all surrogate distributions conform to a single curve, indicating that the sleep-stage
stratification we observe for the real data reflects indeed changes in physiologic coupling
with sleep-stage transitions. In contrast, the same surrogate test applied to traditional cross-correlation
analysis does not show a difference between the rank distributions from surrogate
and real data (Fig. \ref{Rank-CCF}).

\begin{figure}
\includegraphics[scale=0.35]{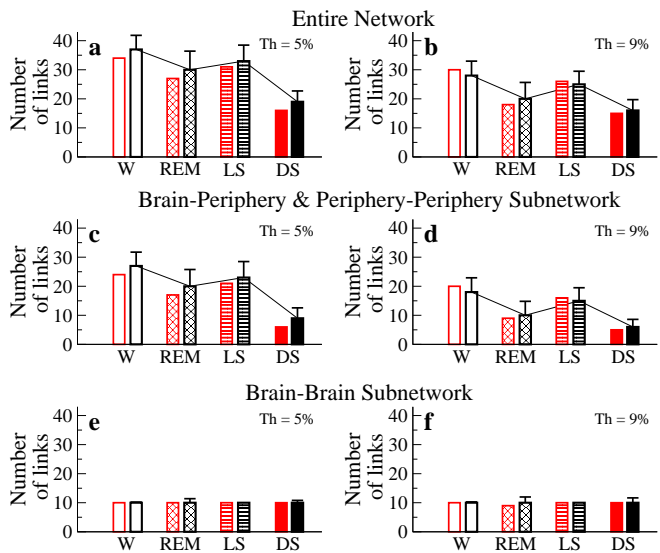}
\caption{{\bf Stability of sleep-stage stratification pattern in network connectivity.}
Group-averaged number of network links for two different thresholds (Th) during wake,
REM, light and deep sleep. Results for threshold of Th = 5\% time delay stability (TDS)  are shown in a, c and e, and
results for threshold of Th = 9\% TDS are shown in b, d and f. The sleep-stage stratification 
pattern observed for the significance threshold of 7\% TDS (shown in Fig. \ref{Histograms})
is preserved also for thresholds of 5\% and 9\% TDS, indicating stability of the results.
Note, that the number of links in the brain-brain subnetwork remains unchanged for 
different sleep stages (e, f), since the strength of all links in this subnetwork is
well above 9\% TDS (Fig. \ref{Histograms}f).}
\label{diff-thresholds}
\end{figure}

We find that the TDS method is better suited than the traditional cross-correlation analysis in identifying
networks of endogenous physiologic interactions. Rank plots obtained from cross-correlation analysis (Fig. \ref{Rank-CCF}) show
that the cross-correlation strength $C_{max}$ (global maximum of the cross-correlation function) is
consistently lower for all links during deep sleep, higher for light sleep and REM and highest during wake --- a stratification
related to the gradual increase in the strength of autocorrelations in the signal output of physiologic 
systems \citep{Bunde-PRL-2000,Schumann-Sleep-2010}, which in turn increases the
degree of cross-correlations \citep{Podobnik-EPJB-2007}. Surrogate tests based on pairs of signals from different
subjects, where the coupling between systems is abolished but physiologic autocorrelations are preserved,
show no statistical difference between the surrogate (open symbols) and original (filled symbols) 
rank distributions of $C_{max}$, suggesting that in this context cross-correlations do not provide physiologically relevant information 
regarding the interaction between systems. Indeed, even for uncoupled systems high autocorrelations in the 
output signals lead to spurious detection of cross-correlations \citep{Podobnik-EPJB-2007}. In contrast, the TDS method is not affected 
by the autocorrelations --- surrogate rank plots for different sleep stages collapse and do not exhibit vertical stratification 
as shown in (Fig. \ref{RankPlots}a).

To test the robustness of the stratification pattern in network topology and connectivity 
across sleep stages (shown in Fig. \ref{TDS-Matrices} and Fig. \ref{Histograms}), we repeat our
analyses for two additional thresholds: 5\% TDS and 9\% TDS. With increasing the threshold 
for TDS from 5\% to 9\% the overall number of links in the network decreases (compare 
Fig. \ref{diff-thresholds}a,c,e with Fig. \ref{diff-thresholds}b,d,f). However, the general sleep-stage
stratification pattern is preserved with highest number of links during light sleep and wake,
lower during REM, and significant reduction in network connectivity during deep sleep (Fig. \ref{diff-thresholds}). The stability
of the observed pattern in network connectivity for a relatively broad range around the significance
threshold of 7\% TDS indicates that the identified network is a robust measure of physiologic interactions. 

\section*{Acknowledgments}
We thank T. Penzel for providing data and helpful comments, and 
A. Y. Schumann for help with data selection, data pre-processing and discussions. We acknowledge support from
NIH Grant 1R01-HL098437, the US-Israel Binational Science Foundation (BSF Grant 2008137), the Office of Naval Research
(ONR Grant 000141010078), the Israel Science Foundation, the European 
Community (projects DAPHNet/FP6 IST 018474-2 and SOCIONICAL/FP7 ICT 231288) and 
the Brigham and Women's Hospital Biomedical Research Institute Fund. 
R.P.B. acknowledges support from the German Academic Exchange Service (DAAD fellowship 
within the Postdoc-Programme).

\section*{Author contributions}
A.B. and R.P.B. contributed equally to this work. A.B., R.P.B., J.W.K., S.H. and P.Ch.I.
designed research. A.B. and R.P.B. wrote the algorithm. A.B., R.P.B. and P.Ch.I. analysed 
the data. R.P.B. and P.Ch.I. wrote the paper with contributions from all.

\section*{Additional information}
Competing financial interests: The authors declare no competing financial interests.

Correspondence and requests for materials should be addressed to P.Ch.I. (email: plamen@buphy.bu.edu or pivanov@partners.org).

\end{document}